\documentclass[11pt,twoside]{article}
\usepackage{fancyhead}
\usepackage{amssymb, amsmath, amsthm}

\newcommand{\beq}[1]{\begin{equation}\label{#1}}
\newcommand{\eeq}{\end{equation}}
\newcommand{\req}[1]{(\ref{#1})}
\newcommand{\bmu}[1]{\begin{multline}\label{#1}}
\newcommand{\emu}{\end{multline}}

\renewcommand{\[}{\left[}
\renewcommand{\]}{\right]}

\newcommand{\eq}{\triangleq}

\newcommand{\A}{\mathcal{A}}

\renewcommand{\S}{\mathcal{S}}

\newcommand{\x}{{\textbf{\textit{x}}}}

\renewcommand{\u}{{\textbf{\textit{u}}}}
\renewcommand{\v}{{\textbf{\textit{v}}}}

\newcommand{\e}{\varepsilon}
\newcommand{\p}{{\textbf{\textit{p}}}}

\newcommand{\EthrU}{\underline{E}_{\text{Thr}}}

\newtheoremstyle{agdTheorem}{\parskip}{\parskip}{\itshape}{\parindent}{\bfseries}{}{0pt}{\thmname{#1}\thmnumber{~#2}.\thmnote{~\textnormal{#3.}}\quad}
\theoremstyle{agdTheorem}
\newtheorem{theorem}{Theorem}

\newtheorem{proposition}{Proposition}

\newtheoremstyle{agdDefinition}{\parskip}{\parskip}{}{\parindent}{\bfseries}{}{0pt}{\thmname{#1}\thmnumber{~#2}.\thmnote{~\textnormal{#3.}}\quad}
\theoremstyle{agdDefinition}
\newtheorem{definition}{Definition}
\newtheorem{remark}{Remark}

\renewcommand{\sectionmark}[1]{}
\renewcommand{\subsectionmark}[1]{}
\begin{document}

\vspace*{5mm}

\noindent
\textbf{\LARGE Threshold Decoding for Disjunctive\\ Group Testing
\footnote{The research is supported
in part by the Russian Foundation for Basic Research
under Grant No. 16-01-00440.}}
\thispagestyle{fancyplain} \setlength\partopsep {0pt} \flushbottom
\date{}

\vspace*{5mm}
\noindent
\textsc{A.G. D'yachkov} \hfill \texttt{agd-msu@yandex.ru} \\
\textsc{I.V. Vorobyev} \hfill \texttt{vorobyev.i.v@yandex.ru} \\
\textsc{N.A. Polyanskii} \hfill \texttt{nikitapolyansky@gmail.com} \\
\textsc{V.Yu. Shchukin} \hfill \texttt{vpike@mail.ru} \\
{\small Lomonosov Moscow State University, Moscow, Russia}

\medskip

\begin{center}
\parbox{11,8cm}{\footnotesize
\textbf{Abstract.}
Let $1 \le s < t$, $N \ge 1$ be integers and a complex electronic
circuit of size $t$ is said to be an $s$-active, $\; s \ll t$,
and can work as a system block if
not more than $s$ elements of the circuit are defective.
Otherwise, the circuit is said to be an $s$-defective and should be replaced
by a similar $s$-active circuit.
Suppose that there exists a possibility to run $N$ non-adaptive group tests
to check the $s$-activity of the circuit.
As usual, we say that a (disjunctive) group test yields the positive response
if the group contains at least one defective element.
Along with the conventional decoding algorithm based on disjunctive $s$-codes,
we consider a threshold decision rule with the minimal possible decoding complexity, which
is based on the simple comparison of a fixed threshold $T$, $1 \le T \le N - 1$,
with the number of positive responses $p$, $0 \le p \le N$.
For the both of decoding algorithms we discuss upper bounds on the $\alpha$-level of significance
of the statistical test for the null hypothesis
$\left\{ H_0 \,:\, \text{the circuit is $s$-active} \right\}$ verse the alternative hypothesis
$\left\{ H_1 \,:\, \text{the circuit is $s$-defective} \right\}$.
}

\end{center}

\baselineskip=0.9\normalbaselineskip

\section{Statement of Problem}
\lhead{}
\rhead{}
\chead[\fancyplain{}{\small\sl\leftmark}]{\fancyplain{}{\small\sl\leftmark}}
\cfoot{}
\markboth{\hspace{-0.2cm}Fifteenth International Workshop on Algebraic and Combinatorial Coding Theory\\ June 18-24, 2016, Albena, Bulgaria \hfill pp. 151--156}{}
\setcounter{page}{151}
\quad
Let $N \ge 2$, $t \ge 2$, $s$ and $T$ be integers, where $1 \le s < t$ and $1 \le T < N$.
The symbol $\eq$ denote the equality by definition, $|A|$ -- the size of the set $A$ and
$[N] \eq \{1, 2, \dots, N\}$ -- the set of integers from $1$ to~$N$.
A binary $(N \times t)$-matrix
\beq{code}
X = \| x_i(j) \|, \quad x_i(j) = 0, 1, \quad i \in [N], \; j \in [t],
\eeq
with $t$ columns (\textit{codewords})
$\x(j) \eq (x_1(j), x_2(j), \dots, x_N(j)$, $j \in [t]$,
and $N$ rows $\x_i \eq (x_i(1), x_i(2) \dots, x_i(t))$, $i \in [N]$,
is called a \textit{binary code of length $N$ and size $t = \lfloor 2^{RN} \rfloor$},
where a fixed parameter $R > 0$ is called a \textit{rate} of the code~$X$.
The number of $1$'s in a binary column $\x = (x_1, \dots, x_N) \in \{0, 1\}^N$, i.e.,
$
|\x| \eq \sum_{i = 1}^N x_i,
$
is called a \textit{weight} of~$\x$. A code $X$ is called
a \textit{constant weight binary code of weight $w$}, $1 \le w < N$, if for any $j \in [t]$, the weight $|\x(j)| = w$.
The conventional symbol $\u \bigvee \v$ will be used to denote the disjunctive (Boolean)
sum of binary columns $\u, \v \in \{0, 1\}^N$.
We say that a column $\u$ \textit{covers}
a column $\v$ ($\u \succeq \v$) if $\u \bigvee \v = \u$.

\subsection{Disjunctive and Threshold Disjunctive Codes}
\label{CombinatorialDefinitions}
\headrulewidth 0pt
\lhead[\fancyplain{}{\thepage}]{\fancyplain{}{\rightmark}}
\rhead[\fancyplain{}{\leftmark}]{\fancyplain{}{\thepage}}
\lfoot{}
\rfoot{}
\chead{}
\cfoot{}
\markboth{\textsl{ACCT2016}}{\textsl{D'yachkov, Vorobyev, Polyanskii, Shchukin}}
\quad
\begin{definition}[\cite{d15}]
\label{defDis}
A code $X$~\req{code} is called a \textit{disjunctive $s$-code},
$s \in [t-1]$, if the disjunctive sum of any $s$-subset
of codewords of $X$ covers those and only those
codewords of $X$ which are the terms of the given disjunctive sum.
\end{definition}

Let $\S$, $\S \subset [t]$, 
be an arbitrary fixed collection
of defective elements of size~$|\S|$. For a binary code $X$ and collection $\S$,
define the binary \textit{response vector} of length $N$, namely:
\beq{rv}
\x(\S) \eq \bigvee_{j \in \S} \, \x(j),\quad \text{if}\quad \S \ne\emptyset \quad
\text{and}\quad  \x(\S) \eq (0, 0, \dots, 0)\quad \text{if}\quad \S = \emptyset.
\eeq
In the classical problem of
{\em non-adaptive group testing}, we describe $N$ tests
as a binary $(N \times t)$-matrix $X = \| x_i(j) \|$,
where a column $\x(j)$ corresponds to the $j$-th element, a row $\x_i$ corresponds to the $i$-th test and
$x_i(j) \eq 1$ if and only if the $j$-th element is included into the $i$-th testing group.
The result of each test equals $1$ if at least one defective element is included into the testing group
and $0$ otherwise, so the column of results is exactly equal to the response vector~$\x(\S)$.
Definition~\ref{defDis} of disjunctive $s$-code $X$
gives the important sufficient condition for the evident identification of any unknown
collection of defective elements $\S$ if the number of defective elements~$|\S| \le  s$.
In this case, the identification of the unknown $\S$ is equivalent to discovery of all codewords of
code~$X$ covered by~$\x(\S)$,
and its complexity is equal to the code size~$t$.
Note that this algorithm also allows us to check $s$-activity of the circuit
defined in the abstract. Moreover, it is easy to prove by contradiction that
every code $X$ which allows to check $s$-activity of the circuit without error is a disjunctive $s$-code.

\begin{definition}
\label{defThr}
Let $s$, $s \in [t-1]$, and $T$, $T \in [N-1]$, be arbitrary fixed integers.
A disjunctive $s$-code $X$ of length $N$ and size $t$ is said to be a 
a disjunctive $s$-code with \textit{threshold} $T$
(or, briefly, \textit{$s^{T}$-code})
if the disjunctive sum of any $\le s$ codewords of $X$ has weight $\le T$ and
the disjunctive sum of any $\ge s+1$ codewords of $X$ has weight $\ge T + 1$.

\end{definition}

Obviously, for any $s$ and $T$, the definition of $s^{T}$-code
gives a sufficient condition for code $X$ applied to
the group testing problem described in the abstract of our paper.
In this case, only on the base of the known {\em number of positive responses} $|\x(\S)|$,
we decide that the controllable circuit identified by
an unknown collection $\S$, $\,\S \subset [t]$,
is $s$-active, i.e., the unknown size $|\S| \le s$ ($s$-defective, i.e., the unknown
size $|\S| \ge s+1$)
if 
$|\x(\S)| \le T$ $\left( |\x(\S)| \ge T+1 \, \right)$.

\begin{remark}
The concept of $s^T$-codes was motivated by
troubleshooting in complex electronic circuits using a non-adaptive identification scheme which was
considered in~\cite{zlm76}.
\end{remark}

\subsection{Hypothesis Test}
\quad
Let a circuit of size $t$ is identified by an unknown collection $\S_{un}$, $\S_{un} \subset [t]$,
of defective elements of an unknown size~$|\S_{un}|$ and
$X$ be a code~\req{code} of size $t$ and length~$N$.
Introduce the null hypothesis
$\left\{ H_0 \,:\, |\S_{un}| \le s \right\}$ (the circuit is $s$-active)
verse the alternative $\left\{ H_1 \,:\, |\S_{un}| \ge s+1 \right\}$ (the circuit is $s$-defective).
In this paper we focus on the testing of statistical hypotheses $H_0$ and~$H_1$.
The similar problem related to constructing of a confidence interval for~$|\S_{un}|$
was considered in~\cite{dam10}, where the authors construct the interval $[\hat s / c \, ; \, \hat s]$,
such that given a {\em random code} $X$, the statistic $\hat s$, i.e., a function 
of the random response vector $\x(\S_{un})$, satisfies the following properties:
$\Pr\{\hat s < |\S_{un}|\}$ is upper bounded by a small parameter $\epsilon \ll 1$
and the expected value of~$\hat s / |\S_{un}|$ is upper bounded by a number~$c > 1$.

For fixed parameters $s$, $1 \le s < t$, and $T$, $1 \le T < N$,
consider the following {\em threshold decision rule}
motivated by Definition~\ref{defThr}, namely:
\beq{dr}
\begin{cases}\quad
\text{accept $\left\{ H_0 \,:\, |\S_{un}| \le s \right\}$} & \text{if $|\x(\S_{un})| \le T$}, \\
\quad
\text{accept $\left\{ H_1 \,:\, |\S_{un}| \ge s+1 \right\}$} & \text{if $|\x(\S_{un})| \ge T+1$}. \\
\end{cases}
\eeq

For the conventional {\em statistical} interpretation of the decision rule~$\req{dr}$,
it is reasonable to assume that
the different collections of defective elements of the same size are {\em equiprobable}. That is why,
we set that the {\em probability distribution} of the random collection $\,\S_{un}$,
$\S_{un} \subset [t]$,
is identified by an unknown probability vector  $\p \eq (p_0, p_1, \dots, p_t)$, $p_k \ge 0$, $k = 0, 1, \dots, t$,
$\sum_{k = 0}^{t} p_k = 1$, as follows:
\beq{pd}
\Pr\{\S_{un} = \S\} \eq \frac{p_{|\S|}}{{t \choose |S|}} \quad \text{for any subset}\quad
\quad \S\subseteq[t].
\eeq

Introduce a {\em maximal error probability} 
of the decision rule~$\req{dr}\;$: 
\beq{maxerror}
\e_s(T, \p, X) \eq \max\, \left\{ \Pr\{ \text{accept } H_1 \big| H_0 \} \, , \, \Pr\{ \text{accept } H_0 \big| H_1 \} \,\right\}
\eeq
where the conditional probabilities in the right-hand side of~\req{maxerror} are identified by~\req{dr}-\req{pd}.
Note that the number $\e_s(T, \p, X) = 0$ if and only if the code $X$ is an $s^T$-code.
Denote by $t_s(N, T)$ the maximal size of $s^{T}$-codes of length~$N$.
For a parameter $\tau$, $0 < \tau < 1$, introduce the rate
of $s^{\lfloor \tau N \rfloor}$-codes:
\beq{Rstau}
R_s(\tau) \eq \varlimsup_{N \to \infty} \frac{\log_2 t_s(N, \lfloor \tau N \rfloor)}{N} \ge 0,
\quad 0 < \tau < 1.
\eeq

\begin{definition}
\label{universalerror}
Let $\tau$, $0 < \tau < 1$, and a parameter $R$, $R > R_s(\tau)$, be fixed.
For the maximal error probability $\e_s(T, \p, X)$, defined by~\req{dr}-\req{maxerror}, consider the
function
\beq{universal-error}
\e^N_s(\tau, R) \eq \, \max\limits_{\p} \,
\left\{ \min\limits_X \, \e_s( \lfloor \tau N \rfloor, \p, X) \, \right\},
\quad
R > R_s(\tau),
\eeq
where the minimum is taken over all codes $X$ of length $N$ and
size $t = \lfloor 2^{RN} \rfloor$ with parameter $R > R_s(\tau)$.
The number $\e^N_s(\tau, R) > 0$ does not depend on
the unknown probability vector $\p$ and can be
called the {\em universal} error probability of the decision rule~$\req{dr}$.
The corresponding {\em error exponent}
\beq{exponent}
E_s(\tau,R) \eq \varlimsup_{N \to \infty} \frac{-\log_2 \e^N_s(\tau , R)}{N},
\quad  s \ge 1,\quad 0 < \tau < 1,\quad R > R_s(\tau)
\eeq
identifies the asymptotic behavior of
{\em $\alpha$-level of significance}
for the decision rule~$\req{dr}$,
i.e.,
\beq{lev-signif}
\alpha \eq \exp_2 \{ -N \, [E_s(\tau, R) + o(1)] \},\quad \text{if}\quad E_s(\tau, R) > 0,\quad N \to \infty.
\eeq
\end{definition}

Along with~\req{dr} we introduce the \textit{disjunctive decision rule} based on the conventional algorithm:
\beq{drDisjunctive}
\begin{cases}\quad
\text{accept $H_0$} & \text{if $\x(\S_{un})$ covers $\le s$ codewords of $X$}, \\
\quad
\text{accept $H_1$} & \text{if $\x(\S_{un})$ covers $\ge s+1$ codewords of $X$}. \\
\end{cases}
\eeq
For a fixed code rate $R$, $R > 0$, the error exponent for disjunctive decision rule~\req{drDisjunctive}
$E_s(R)$ is defined similarly to~\req{maxerror}-\req{exponent}.
The function $E_s(R)$ was firstly introduced in our paper~\cite{d15}, where
we proved
\smallskip

\begin{theorem}[\cite{d15}]
\label{thCapacityUpper}
The function $E_s(R) = 0$ if $R \ge 1/s$.
\end{theorem}

\section{Lower Bounds on Error Exponents}
\label{secBounds}
\quad
In this Section, we formulate and compare the random coding lower
bounds for the both of error exponents $E_s(R)$ and~$E_s(\tau,R)$.
These bounds were proved applying the random coding method based
on the ensemble of constant-weight codes.
A parameter $Q$ in formulations of theorems~\ref{thCapacityLower}-\ref{thRandomEthr}
means the relative weight of codewords of constant-weight codes.
Introduce the standard notations
$$
h(Q) \eq -Q \log_2 Q - (1 - Q) \log_2 [1 - Q],\quad
[x]^+ \eq \max\{x, 0\}.
$$
In~\cite{d15}, we established
\begin{theorem}[\cite{d15}]
\label{thCapacityLower}
The error exponent $E_s(R) \ge \underline{E}_s(R)$
where the random coding lower bound
$$
\underline{E}_s(R) \eq \, \max_{0 < Q < 1} \, \min_{Q \le q < \min\{1, sQ\}} \left\{ \A(s, Q, q)
+ \[h(Q) - q h(Q / q) - R\]^+\right\},
$$
$$
\A(s, Q, q) \eq (1-q) \log_2(1-q) + q \log_2 \[ \frac{Qy^s}{1-y} \]
+ sQ \log_2 \frac{1-y}{y} + sh(Q),
$$
and $y$ 
is the unique root of the equation
$$
q = Q \frac{1 - y^s}{1 - y}, \quad 0 < y < 1.
$$
In addition, as $s \to \infty$ and $R \le \frac{\ln 2}{s}(1 + o(1))$,  the lower bound
$\underline{E}_s(R)>0$.
\end{theorem}


\begin{theorem}
\label{thRandomEthr}
\textnormal{\textbf{1.}}
The error exponent $E_s(\tau, R) \ge \underline{E}_s(\tau, R)$
where the random coding bound $\underline{E}_s(\tau, R)$ does not depend
on $R > 0$ and has the form:
$$
\underline{E}_s(\tau,R) \eq  \max_{1 - (1 - \tau)^{1 / (s + 1)} < Q < 1 - (1 - \tau)^{1 / s}}
\min \left\{ \A'(s, Q, \tau), \A(s+1, Q, \tau) \right\},
$$
$$
\A'(s, Q, q) \eq
\begin{cases}
\A(s, Q, q), \quad &\text{ if } Q \le q \le sQ,\\
\infty, \quad &\text{ otherwise}.
\end{cases}
$$
\textnormal{\textbf{2.}}
As $s \to \infty$ the optimal value of $\underline{E}_s(\tau, R)$
satisfies the inequality:
$$
\EthrU(s) \eq \max_{0 < \tau < 1} \underline{E}_s(\tau,R) \ge \frac{\log_2 e}{4s^2} (1 + o(1)), \quad s \to \infty.
$$
\end{theorem}

It is possible to use decision rule~\req{dr} with any value
of parameter $T$. The numerical values of the optimal error exponent $\EthrU(s)$
along with the corresponding optimal values of threshold parameter $\tau$
and the constant-weight code ensemble parameter $Q$
are presented in Table~1. Besides in Table~1
the values of $\underline{E}_s(0)$ and
$R_{\text{cr}} \eq \sup\{ R : \underline{E}_s(R) > \EthrU(s) \}$ are shown.
Theorems~\ref{thCapacityUpper}-\ref{thRandomEthr} show that, for large values of $R$,
the threshold decision rule~\req{dr}
has an advantage over the disjunctive decision rule~\req{drDisjunctive} as $N \to \infty$.

\begin{table}[ht]
\label{EthrU}
\caption{The numerical values of $\EthrU(s)$}
\begin{center}
\begin{tabular}{|c|ccccccc|}
\hline
$s$ & $2$ & $3$ & $4$ & $5$ & $6$ & $7$ & $8$\\
\hline
$\EthrU(s)$ & $0.1380$ & $0.0570$ & $0.0311$ & $0.0196$ & $0.0135$ & $0.0098$ & $0.0075$\\
\hline
$\tau$ & $0.2065$ & $0.1365$ & $0.1021$ & $0.0816$ & $0.0679$ & $0.0582$ & $0.0509$\\
\hline
$Q$ & $0.1033$ & $0.0455$ & $0.0255$ & $0.0163$ & $0.0113$ & $0.0083$ & $0.0064$\\
\hline
$\underline{E}_s(0)$ & $0.3651$ & $0.2362$ & $0.1754$ & $0.1397$ & $0.1161$ & $0.0994$ & $0.0869$\\
\hline
$R_{\text{cr}}$ & $0.2271$ & $0.1792$ & $0.1443$ & $0.1201$ & $0.1027$ & $0.0896$ & $0.0794$\\
\hline
\end{tabular}
\end{center}
\end{table}

\section{Simulation for finite code parameters}
\label{secSimulation}
For finite $N$ and $t$, we carried out a simulation as follows.
The probability distribution vector $\p$~\req{pd} is defined by
$$
p_k \eq {t \choose k} p^k (1 - p)^{t - k}, \quad p \eq \frac{s + 1/2}{t}, \quad 0 \le k \le t,
$$
i.e. the number of defective elements $|\S_{un}|$ has the binomial distribution with the mean value~$s + 1/2$.
A code $X$ is generated randomly from the ensemble of constant-weight codes, i.e.
for some weight parameter $w$, every codeword of $X$ is chosen independently and equiprobably
from the set of all ${t \choose w}$ codewords. For every weight $w$ and every decision rule,
we repeat generation $1000$ times and choose the code
with minimal error probability. Note that for disjunctive decision rule $\Pr\{\text{accept } H_0 | H_1\} = 0$.
In Table~2 the best values of maximal error probability for fixed
parameters $s$, $t$ and $N$ are shown in bold.

\begin{table}[ht]
\label{sim15}
\caption{Results of simulation}
\begin{center}
\begin{tabular}{|c|cccc|cc|}
\hline
& \multicolumn{4}{|c|}{Threshold decision rule}
& \multicolumn{2}{|c|}{Disjunctive decision rule}\\
\hline
$N$ & $\Pr\{\text{acc. } H_1 | H_0 \}$ & $\Pr\{\text{acc. } H_0 | H_1 \}$ & $w$ & $T$
& $\Pr\{\text{acc. } H_1 | H_0 \}$ & $w$\\
\hline
\multicolumn{7}{c}{$s = 2, \quad t = 15$}\\
\hline
$5$ & $\mathbf{0.1366}$ & $0.1355$ & $2$ & $3$ & $0.4780$ & $2$\\
\hline
$8$ & $0.0732$ & $\mathbf{0.0824}$ & $3$ & $5$ & $0.3610$ & $2$\\
\hline
$10$ & $0$ & $\mathbf{0.0744}$ & $1$ & $2$ & $0.2390$ & $3$\\
\hline
$12$ & $0$ & $\mathbf{0.0440}$ & $1$ & $2$ & $0.1220$ & $3$\\
\hline
$14$ & $0$ & $\mathbf{0.0349}$ & $2$ & $4$ & $0.0537$ & $3$\\
\hline
$15$ & $0$ & $0.0258$ & $2$ & $4$ & $\mathbf{0.0195}$ & $3$\\
\hline
\multicolumn{7}{c}{$s = 2, \quad t = 20$}\\
\hline
$5$ & $\mathbf{0.1398}$ & $0.1365$ & $2$ & $3$ & $0.5356$ & $2$\\
\hline
$8$ & $\mathbf{0.0897}$ & $0.0890$ & $3$ & $5$ & $0.4169$ & $2$\\
\hline
$10$ & $\mathbf{0.0897}$ & $0.0858$ & $3$ & $5$ & $0.3008$ & $3$\\
\hline
$12$ & $\mathbf{0.0580}$ & $0.0576$ & $4$ & $7$ & $0.1979$ & $3$\\
\hline
$15$ & $0$ & $\mathbf{0.0324}$ & $2$ & $4$ & $0.0792$ & $4$\\
\hline
\end{tabular}
\end{center}
\end{table}


\begin{thebibliography}{99}
\markboth{\textsl{ACCT2016}}{\textsl{D'yachkov, Vorobyev, Polyanskii, Shchukin}}

\bibitem{d15}
\textit{D'yachkov A.G.}, \textit{Vorobyev I.V.}, \textit{Polyanskii N.A.}, \textit{Shchukin V.Yu.},
Almost Disjunctive List-Decoding Codes,
\textit{Problems of Information Transmission}, vol. 51, no. 2, pp.~110-131, 2015.

\bibitem{zlm76}	
\textit{Zubashich V.F.}, \textit{Lysyansky A.V.}, \textit{Malyutov M.B.},
Block–randomized distributed
trouble–shooting construction in large circuits with redundancy.
\textit{Izvestia of the USSR Acad. of Sci., Technical Cybernetics}, vol.~6, 1976.

\bibitem{dam10}
\textit{Damaschke P.}, \textit{Muhammad A.S.},
Competitive group testing and learning hidden vertex covers with minimum adaptivity,
\textit{Discrete Math. Algorithm. Appl.}, vol. 2, no. 3, pp. 291-311, 2010.

\end{thebibliography}
\end{document}